# EPICS Software Development for SNS VME Based Timing and RTDL System*

J. Y. Tang[†], H. Hartmann, L. Hoff, T. Kerner, B. Oerter and J. Smith
Brookhaven National Laboratory, Upton, NY, USA
P. McGehee and P. Stein, Los Alamos National Laboratory, Los Alamos, NM, USA

Abstract

The Spallation Neutron Source (SNS) timing and Real Time Data Link (RTDL) systems are being designed and developed at Brookhaven National Laboratory (BNL), Los Alamos National Laboratory (LANL) and other SNS collaborating labs [1]. The VME-based SNS timing system is comprised of a phase locked loop (PLL) to generate a 120 Hz power line locked clock, a master timing system including up to 4 16-event input modules (V101) and a timing system encoder module (V123S), distribution system to all equipment locations, and a timing system receiver module (V124S). The VME-based SNS RTDL system includes an encoder module (V105), up to 10 8-inputs modules (V206) and an RTDL receiver module. The EPICS [2] software support for the timing and RTDL master VME boards has been developed at BNL while the software and hardware of the SNS utility module are undertaken at LANL. The software is currently under the beta test at the SNS collaborating labs. This paper describes the design and the implementation of the software.

## 1 INTRODUCTION

The design of the SNS beam synchronous timing system is based on the existing model for the Brookhaven National Laboratory's Relativistic Heavy Ion Collider (RHIC). Adapting the RHIC system for SNS reduces the development time and takes advantage of proven technology reducing the engineering risk. Modifications to the RHIC system to meet SNS requirements would be minor. For the same reasons, the most of related RHIC and APS timing control software has been ported and re-engineered. The SNS timing software development is currently being developed on the SNS timing system test stand at BNL. Its configuration is shown in Figure 1. The software will be installed and used for the SNS control system commissioning at Oak Ridge National Laboratory.

## 2 SNS TIMING SYSTEM HARDWARE

### 2.1 Beam Synchronous Event Encoder System

The design of SNS beam synchronous event encoder is adapted from the RHIC system [3] and it is contained in a single VME chassis utilizing five dedicated slots, one for the V123S module and four for V101 modules. The slots are fixed by the dedicated P2 backplane. The beam synchronous event link carrier varies around 16.92 MHz, derived from 16 x the SNS ring LLRF frequency. The source of this RF clock is created by direct digital synthesizer (DDS) and a power line tracking (PLL) VME board. The beam synchronous event link is initiated and fanned out as differential PECL, and distributed by fiber optics.

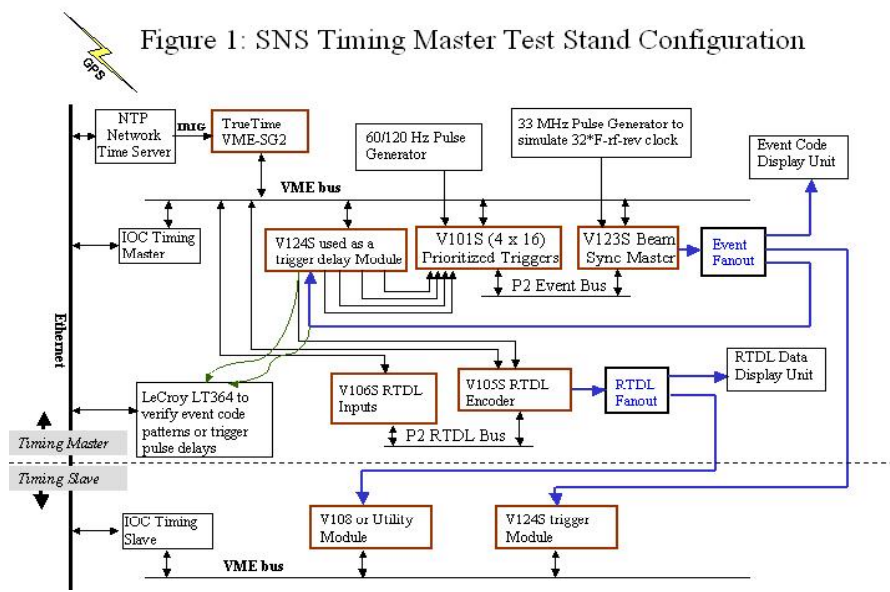

Figure 1: SNS Timing Master Test Stand Configuration

The purpose of the SNS beam synchronous event system is:

- Transmit the 600 Hz Line sync Event to establish spill and fill cycle delineation points accurate to 1 ns
- Synchronize the above signals to the 1 MHz ring rotation tick
- Synchronize the neutron choppers to the ring extraction time
- Synchronize the linac beam chopper injection pulse to the accumulator ring phase
- Provide beam synchronous events to the instrumentation triggers, beam position monitors, and profilers
- Track variations in injection energy from 842 MeV to 1.3 GeV or B field corrections requiring rotation frequency adjustments

## 2.2 The Beam Synchronous Trigger Module

The beam synchronous trigger module (V124S) is a general-purpose VME bus controlled module and is designed to provide clocks and triggers for data acquisition systems and experiments. The V124S module provides eight identical channels that can be configured independently and a buffered recovered RF Clock output. The design of this module is based on the RHIC's version [4]. The primary function of this module is the trigger delay function.

## 2.3 SNS Real Time Data Link System

The SNS RTDL system is comprised of two module types: the Encoder Module (V105) and the Input Module (V206). There is only one V105 module, but many (up to 10) Input Modules. The design of the SNS RTDL system is based on the RHIC's version [5]. The RTDL system distributes to all locations around the SNS machine parameters of general interest to accelerator systems and users. The RTDL machine parameter frames are transmitted on a serial self-clocking link using a modified Manchester code. The transmission rate is 10Mb/s and about 4.3 us are required to transmit each parameter data frame. A typical application of using the system is to distribute a GPS time stamp to all the IOCs around the SNS machine.

## 2.4 The Utility Module

The SNS utility module is a multi-purpose circuit board used to provide an Input/Output Controller (IOC) with interfaces to SNS event and RTDL links, input/output capability, and chassis environment information. The utility module, designed to reside in a VME chassis, occupies one VME board slot. It has the following main features and functions:

- SNS event link decoder and event link filter used to generate VME interrupts
- SNS RTDL utility link receiver and data frame buffer
- RTDL utility link error monitor
- Status indication for event link, SNS utility link and board initialization
- Power supply, fan and temperature monitoring for the VME chassis
- Interrupt generation from failed links, power supplies, fans and over temperature
- Remote reset of VME chassis via the RTDL utility link
- VME interrupts from two external signal sources

# 3 SOFTWARE SUPPORT

## 3.1 VxWorks Driver Development

The VxWorks drivers for the SNS timing hardware provide EPICS device support layer or high-level application software the access to the hardware registers and the capability of the hardware module configuration. The driver development strategy taken was Wind River VxWorks driver compliant: top-down design and bottom-up implementation and testing. The original RHIC drivers were used as template files for the SNS timing drivers.

*bsyncDrv()* – This driver provides an interface to the v123s, the SNS beam-synchronous encoder module and the 16-channel v101 event code input module. 20 routines implemented in this driver are accessible through the VxWorks I/O system for I/O access to specific registers. The interrupt service routine implemented accumulates input errors from various sources.

*rtdlDrv()* – This driver controls the SNS RTDL system. 25 routines implemented in this driver are accessible through the VxWorks I/O system for controls of the RTDL encoder, v105 module and the input module, v106 via their registers. In addition to the IOCTL routines, the read and write functions of the driver are provided for reading or writing from the parameter ID table within the encoder module. The interrupts are generated when the encoder module does not get a response from a particular input channel, as well as when no event line carrier has been detected for 1 us. The Parameter ID on the encoder module is read by the interrupt handler when a NO RESPONSE interrupt occurs in order to decipher which channel is at fault. The interrupts are generated on the input module when the carrier on the local serial link has not been detected for 2 us, when a frame has been received with a data integrity (CRC) error, and when a frame error occurs.

*bsTrigDrv()* – This driver provides an interface to the v124s 8-channel, beam-synchronous trigger module. The driver allows configuring each of the 8-channel individually. 38 IOCTL routines implemented in this driver are accessible through the I/O system for configuring the module.

*utilityDrv()* – This driver consists of read and write access to specific registers in addition to handling of three distinct interrupt sources. With the exception of the RTDL frames all registers are accessed as 8-bit quantities. The driver reads error counts and status for the event and RTDL links, the RTDL frames, environmental parameters associated with the power supply and temperature control, and the hardware selected remote reset address. Write access to the utility board allows control of which event codes generate local interrupts, temperature high limit set point, and various initialization functions. Interrupt service routines are attached to interrupts stemming from the Event Link, from the monitoring of the board environment, and from external sources.

*3.2 EPICS Device Support Layer*

EPICS device support layer for the SNS timing system hardware serves the interface between the VxWorks drivers and EPICS control system. After careful review of APS' timing system EPICS support, instead of creating a new set of record types, a group of record templates with a new set of device support routines has been designed and implemented for the SNS timing system EPICS support. This choice has been found to be more flexible for a prototyping hardware system and easier for the future EPICS upgrading effort. Three major function components of the SNS timing system EPICS support are the basic I/O interface, synchronized time stamp support and triggering of EPICS database processing in response to event link activity.

*The basic I/O interface* - this provides a mapping between the EPICS database records and the device driver I/O functions over the board registers. The board configuration can be accomplished via EPICS database records.

*The synchronized time stamp support* – this is accomplished by using the synchronized time stamp support interface originally defined at APS and makes use of three RTDL parameters updated at 60 Hz by a RTDL master. To support this feature, the device support for the utility module enables the interrupt generated by the "RTDL Valid" event at initialization.

*The triggering of EPICS database processing in response to the event link activity* - The device support layer allows the application developer to define records that will be processed when a specific combination of an event and type of beam pulse profile occurs.

## 3   SUMMARY

Both hardware and software for the SNS timing system are currently in beta test at the SNS partner labs. The first application of the software will be conducted at LBNL for SNS Front-End diagnostic system around the end of this year. On-going effort on hardware and software will be continued until the commissioning of the SNS at ORNL in 2004 or 2005.

## 5 ACKNOWLEGEMENTS

The authors wish to thank Marty Kraimer of APS for the meaningful discussions on using record templates vs. creating new record types.